\documentclass[runningheads]{llncs}
\usepackage{graphicx}
\usepackage{makecell}  

\usepackage[ruled,linesnumbered]{algorithm2e} 
\let\oldnl\nl
\newcommand{\nonl}{\renewcommand{\nl}{\let\nl\oldnl}}

\usepackage{amsmath}  

\begin{document}
\title{Efficient Wait-Free Linearizable Implementations of Approximate Bounded Counters Using Read-Write Registers\thanks{This work has been partially supported by the ANR projects SKYDATA (ANR-22-CE25-0008), TEMPOGRAL (ANR-22-CE48-0001) and CNRS-INS2I, Projet RELAX 2022.}}
\titlerunning{Implementations of Approximate Bounded Counters Using Registers}
%
\author{Colette Johnen\inst{1}\orcidID{0000-0001-7170-4521} \and
Adnane Khattabi\inst{2} \and
Alessia Milani\inst{3} \and
Jennifer L.\ Welch\inst{4}\orcidID{0000-0001-7164-1436}}
\authorrunning{C. Johnen et al.}
%
\institute{
   LaBRI, Universit\'{e} de Bordeaux, Bordeaux, France \\
   \email{johnen@labri.fr}
\and
   ESI Group, Lyon, France \\
   \email{adnane.khattabi@esi-group.com}
\and
   LIS, Aix-Marseille Universit\'{e}, Marseille, France \\
   \email{alessia.milani@univ-amu.fr}
\and
   Texas A\&M University, College Station, TX 77843, USA \\
   \email{welch@cse.tamu.edu}
}
\maketitle              
\begin{abstract}   
Relaxing the sequential specification of a shared object is a way to obtain an implementation with better performance compared to implementing the original specification. We apply this approach to the {\em Counter} object, under the assumption that the number of times the Counter is incremented in any execution is at most a known bound $m$. We consider the $k$-{\em multiplicative-accurate} Counter object, where each read operation returns an approximate value that is within a multiplicative factor $k$ of the accurate value. More specifically, a read is allowed to return an approximate value $x$ of the number $v$ of increments previously applied to the counter such that $v/k \le x \le vk$. We present three algorithms to implement this object in a wait-free linearizable manner in the shared memory model using read-write registers. All the algorithms have read operations whose worst-case step complexity improves exponentially on that for an {\em exact} $m$-bounded counter (which in turn improves exponentially on that for an exact {\em unbounded} counter). Two of the algorithms have read step complexity that is asymptotically optimal. The algorithms differ in their requirements on $k$, step complexity of the increment operation, and space complexity. 

\end{abstract}

\section{Introduction}

Finding efficient ways to implement linearizable shared objects out of
other shared objects in crash-prone asynchronous distributed systems
is central to concurrent programming.
In this paper, we focus on implementing the {\it Counter} object, which provides
an Increment operation that increases the value of the counter by one,
and a Read operation that returns the current value of the counter.
Counters are a fundamental data structure for many applications,
ranging from multicore architectures to web-based e-commerce.

A general result by Jayanti, Tan and Toueg~\cite{JayantiTT2000}
implies the discouraging result that any implementation of a Counter
using ``historyless'' objects---those whose modifying operations
over-write each other---has an execution in which some operation takes
$\Omega(n)$ steps on the building block objects, where $n$ is the
number of processes in the system.  As noted in \cite{AspnesAC2012}, this 
lower bound is tight, as
a Counter can be implemented with an atomic snapshot object, which in
turn can be implemented using read-write registers with linear step
complexity~\cite{InoueC94}.

There are several ways one could attempt to circumvent this linear
lower bound on the worst-case step complexity.  
The proof in~\cite{JayantiTT2000} constructs a slow execution in which
a very large number of operations are performed on the implemented object.
A natural restriction is to consider the case when the number of operations,
especially increments, is {\it bounded}, say at most $m$.
This approach is taken by Aspnes, Attiya and Censor-Hillel \cite{AspnesAC2012},
resulting in an algorithm for an exact counter whose worst-case step
complexity is $O(\log m)$ for the Read operation and $O(\log n \cdot \log m)$ 
for the Increment operation\footnote{$\log n$ means $\log_2 n$; 
any other base of a logarithm is explicitly given.
Since Read and Increment can be implemented in $O(n)$ steps, technically each
asymptotic bound of the form $f(k,m,n)$ for some function $f$ should be $\min\{n,f(k,m,n)\}$.  
To simplify the expressions, we implicitly assume that $f(k,m,n)$ is $o(n)$ so that
we can drop the ``minimum of $n$'' part of the expression.}.
The step complexity for Read is tight, as shown by an $\Omega(\log m)$ lower
bound in \cite{AspnesAC2012,AspnesCAH16}.

Another approach is to consider the {\it amortized} step complexity
instead of worst case.  It could be that in any execution, most of the
operations are fast, while only a few are slow.  Baig, Hendler, Milani
and Travers \cite{BaigHMT23} present an exact unbounded Counter
implementation that has $O(\log^2 n)$ amortized step complexity.  
This performance is close to tight, thanks to an $\Omega(\log n)$ lower bound
in \cite{BaigHMT23}, which is based on a result in~\cite{AttiyaH10}.

Finally, the semantics of the Counter object being implemented could
be relaxed so that the value returned by a Read is not necessarily
exactly the number of preceding Increments.  Approximate counting has many applications (e.g.,~\cite{AspnesCAH16,AfekKKMT14}); approximate {\em probabilistic} counting has been studied extensively both in the sequential setting (e.g.,~\cite{Morris78a,Flajolet85}) and the concurrent setting (e.g.,~\cite{AspnesC10,BenderG11}).  We focus on the {\em deterministic} situation.
A Counter is said to be
$k$-{\it multiplicative-accurate} if, informally speaking, each Read
returns a value that is within a factor of $k$ of the Counter value.
This approximation is exploited by Hendler, Khattabi,
Milani and Travers \cite{HendlerKMT21} to improve the {\it amortized} 
step complexity of an unbounded counter.  
They achieve $O(1)$ amortized step complexity 
in any execution if 
$k \ge n$, while
for certain long executions the constant amortized step complexity is achieved for $k \ge \sqrt{n}$ \cite{Khattabi-thesis};
if $k < \sqrt{n}$, they show a lower bound of $\Omega(\log \frac{n}{k^2})$ on the amortized step complexity. 
The constant step complexity upper bound does not contradict the $\Omega(\log n)$ lower bound in \cite{BaigHMT23} since
the lower bound is for exact counters, indicating the performance
benefit resulting from the approximation.

In this paper, we present three implementations of a
$k$-{\it multiplicative-accurate} $m$-{\it bounded} Counter
and analyze their {\it worst-case} step complexities.
All of our algorithms use only read-write registers.  
All our algorithms have $O(\log \log m)$ or smaller Read complexity,
which is an exponential improvement on the $O(\log m)$ bound in \cite{AspnesAC2012},
thanks to the approximation.
Our results are incomparable to those in
\cite{BaigHMT23,HendlerKMT21} 
since we consider {\em worst-case} complexity of {\em bounded} Counters  
instead of {\em amortized} complexity of {\em unbounded} Counters.

We first present a simple Counter implementation in which both Read and Increment
have $O(\log \log m)$ step complexity, as long as $k$ is a real number with $k \ge \sqrt{2n}$.
(See Section~\ref{sec:kAtLeastSquareRoot}.)
The algorithm uses a shared max-register object
that is bounded by $\lceil \log m \rceil$
in which processes store the logarithm of the number of Increments
that they know about so far.
When a process executes an Increment, it keeps track in a local variable of the
number of Increments invoked at it; every time the number has doubled, it writes
a new value to the max-register, which is one larger than the previous value it wrote.
The Read operation reads the max-register and returns $k$ times $2^r$, where
$r$ is the value read from the max-register.
By using the max-register implementation in~\cite{AspnesAC2012}, we obtain
the claimed step complexity.
The idea of waiting to expose the number of Increments until a power has been reached and then writing the logarithm of the number is taken from an algorithm in~\cite{HendlerKMT21}; however, that algorithm waits for powers of $k$ and does not have good complexity in  executions with few increments.
Our innovations are to wait for powers of 2 instead of $k$, and to carefully control the number of increments exposed together, as well as the evolution of this value.

Our second result is a Counter implementation in which the step complexity of Read
is $O(\log \log_k m)$ and that of Increment is $O(\max\{\log n \cdot
\log(\frac{k n}{k-1}), \log \log_k m\})$, for any real number $k$ with $k > 1$.
(See Section~\ref{sec:kGreaterThanOne}.)
The Read complexity is asymptotically optimal due to a lower bound of
$\Omega(\log \log_k m)$ in \cite{HendlerKMT21}.
Like the first algorithm, this one uses a shared max-register.  
To track the number of Increments more accurately while keeping the fast Read operation,
each Increment could increment an exact counter, then read the exact counter and write the logarithm of the value read to the max-register.  However, the exact counter is bounded by $m$, and when implemented with registers using
the algorithm in~\cite{AspnesAC2012} results in an Increment step complexity of $\Omega(\log n \cdot \log m)$.
To reduce the Increment step complexity in the common case when $m$ is much larger than $n$, our algorithm
uses an array of smaller exact counter objects, which we call
``buckets''.  There are $\left\lceil (k-1)\frac{m}{n} \right\rceil$ buckets, with the maximum value stored in a bucket being $\left\lceil \left(\frac{k}{k-1}\right)n \right\rceil$.
Using buckets reduces the step complexity, with the tradeoff that a process cannot always decide if an increment in a bucket has an effect (i.e., it was stored).
By using the exact bounded counter and max-register implementations in \cite{AspnesAC2012}, we obtain the claimed step complexity.

The third algorithm works for any integer $k$ with $k \ge 2$. 
It combines the techniques of the two first algorithms:  exposing increments in batches and using buckets.
It has better space complexity than, and the same Read step complexity as, the second algorithm, but its Increment step complexity is worse when $k$ is super-constant with respect to $n$. (See Section~\ref{sec:kAtLeastTwo}.) 
In more detail, the Increment step complexity is $O(\max\{\log n \cdot \log n, \log \log_k m\})$, the number of buckets is $\left\lceil \log_k \frac{m}{n} \right\rceil$ and the maximum value stored in a bucket is $4n$.

A key challenge for both our second and third algorithms was to prove linearizability.  The definitions of the linearizations are subtle and take into account interactions between multiple operations.

It is easy to see that all our algorithms are wait-free as they have no loops or waiting statements.
Our results are summarized in Table~\ref{table:3-algs}. 

\begin{table}[ht]
\caption{Comparison of our $k$-multiplicative-accurate $m$-bounded Counter implementations}
\centering
\begin{tabular}{|c||c|c|c|}
\hline
     & Algorithm~\ref{alg:kAtLeastSquareRoot} 
     & Algorithm~\ref{alg:kLargerThanOne} 
     & Algorithm~\ref{alg:kAtLeastTwo} 
     \\
\hline
\hline
    $k$ 
    & $k \ge \sqrt{2n}$ 
    & $k > 1$ 
    & $k \ge 2$ 
    \\               
\hline
    \makecell{Read step \\ complexity} 
    & $O(\log \log m)$ 
    & \makecell{$O(\log \log_k m)$ \\ optimal} 
    & \makecell{$O(\log \log_k m)$ \\ optimal} 
    \\     
\hline
    \makecell{Increment step \\ complexity} 
    & $O(\log \log m)$ 
    & \makecell{$O(\max\{\log n \cdot \log \frac{k}{k-1}n,$ \\ $ ~~~~~~~~\log \log_k m\})$} 
    & $O(\max\{\log^2 n, \log \log_k m \})$ 
    \\
\hline
    \makecell{space \\ complexity}
    & 1 max-register 
    & \makecell{1 max-register \\ $\left\lceil \frac{(k-1)m}{n} \right\rceil$ exact counters, \\ each $\left\lceil \frac{kn}{k-1}\right\rceil$-bounded} 
    & \makecell{1 max-register \\ $\left\lceil \log_k \frac{m}{n} \right\rceil$ exact counters, \\ each $4n$-bounded} 
    \\ 
\hline
\end{tabular}
\label{table:3-algs}
\end{table}

\section{Preliminaries}
\label{section:model}

{\it Overview.}
We consider an asynchronous shared memory system in which a set $\cal
P$ of $n$ crash-prone processes communicate by applying operations on
shared objects.  The objects are {\it linearizable}, which means that
each operation appears to occur instantaneously at some point between
its invocation and response and it conforms to the sequential
specification of the object~\cite{HerlihyW90}.  Our ultimate goal is
for the system to implement a linearizable $k$-multiplicative-accurate
counter by communicating using linearizable read-write registers.
However, for convenience, our algorithms are described using
linearizable max-register objects and linearizable exact counter
objects, which in turn can be implemented using linearizable read-write
registers.  When proving the correctness of our algorithms, we assume
that the operations on the max-registers and exact counters are
instantaneous, but when analyzing the step complexity of our
algorithms, we take into account the specific algorithms used to
implement the max-registers and exact counters out of registers.

{\it Sequential Specifications of Objects.}
We next give the sequential specifications of the objects under consideration.
\begin{itemize}
\item A {\it read/write register} has operations Read and Write; in every
  sequence of operations, each Read returns the value of the latest preceding
  Write (or the initial value if there is none).

\item A {\it max-register} has operations MaxRead and MaxWrite; in every
  sequence of operations, each MaxRead returns the largest value among all the
  preceding MaxWrites (or 0 if there is none).
  A max-register is $h$-{\it bounded} if attention is restricted to sequences of
  operations in which the largest value of any MaxWrite is $h$.

\item An {\it exact Counter} has operations Read and Increment;
  in every sequence of operations, each Read returns the number of
  preceding Increments.  

\item A $k$-{\it multiplicative-accurate Counter} has operations Read and
  Increment; in every sequence of operations, each Read
  returns a value $x$ such that $v/k \le x \le kv$, where $v$ is the number of preceding
  Increments.

\item A Counter (either exact or multiplicative-accurate) is $h$-{\it bounded} if attention
  is restricted to sequences of operations in which at most $h$ Increments occur.
\end{itemize}

{\it Executions of Implementations.}
An {\it implementation} of a shared object provides a specific data representation for
the object from base objects, 
each of which is assigned an initial value.
The implementation also provides sequential algorithms for
each process in $\cal P$ that are executed when operations on the implemented
object (a $k$-multiplicative-accurate $m$-bounded Counter in our case) is invoked;
these algorithms involve local computation and operations on the base objects.

Each {\it step} of a process contains at most one invocation of an operation of
the implemented object, at most one operation on a base object, and at most
one response for an operation of the implemented object.  A step can also contain
local computation by the process.

An {\it execution} of an implementation of a shared object is a
possibly-infinite sequence of process steps such that the subsequence
of the execution consisting of all the steps by a single process is
{\it well-formed}, meaning that the first step is an invocation,
invocation and responses alternate, and the steps between an
invocation and its following response are defined by the algorithm
provided by the implementation.  Since we put no constraints on the
number of steps between consecutive steps of a process or between
steps of different processes, we have modeled an {\it asynchronous}
system.

{\it Wait-freedom.}  We desire algorithms that can tolerate any number
of crash failures.  This property is captured by the notion of {\it
wait-freedom}~\cite{Herlihy91}: in every execution, if a process takes
an infinite number of steps, then the process executes an infinite
number of operations on the implemented object.  In other words, each
process completes an operation if it performs a sufficiently large
number of steps, regardless of how the other processes' steps are
scheduled.

The rest of this section is devoted to a lemma showing that a generic way of ordering the operations in an execution ensures the relative order of non-overlapping operations.  
The lemma is independent of the semantics of the object being implemented.
This behavior is part of what is needed to prove linearizability.  
The other part, showing that the sequential specification is respected, of course depends on the specific object being implemented and is not addressed by this lemma.
The purpose of extracting this observation as a stand-alone lemma is that essentially the same argument is used in the linearizability proofs for Algorithms~\ref{alg:kLargerThanOne} and~\ref{alg:kAtLeastTwo}.

Partition all the complete, and any subset of the incomplete, operations in a concurrent execution $E$ into two sets, $A$ and $B$.  Create a total order $L$ of the operations as follows:

\begin{enumerate}
\item Choose any point inside each operation in $A$ and order these
operations in $L$ according to the chosen points.

\item Consider the operations in $B$ in increasing order of when they
start in $E$.  Let $op$ be the next operation under consideration.
Let $op'$ be the earliest operation already in $L$ that begins in $E$
after $op$ ends in $E$.  Place $op$ immediately before $op'$ in $L$.
\end{enumerate}

\begin{lemma}
\label{lem:generic-ordering}
$L$ respects the order of non-overlapping operations in $E$.
\end{lemma}

\begin{proof}
Let $op_1$ and $op_2$ be two operations such that $op_1$ ends before $op_2$
begins in $E$.  We will show that $op_1$ precedes $op_2$ in $L$.

{\em Case 1:} Both $op_1$ and $op_2$ are in $A$.  Then they are
ordered in $L$ according to the linearization points inside their
intervals of execution and thus $op_1$ precedes $op_2$ in $L$.

{\em Case 2:} Both $op_1$ and $op_2$ are in $B$.  Let $op_1'$ (resp.,
$op_2'$) be the earliest operation already in $L$ when $op_1$ (resp.,
$op_2$) is being placed that starts after $op_1$ (resp., $op_2$) ends
in $E$.  Note that neither $op_1'$ nor $op_2'$ is in $B$,
since operations in $B$ are considered for placement in the order in
which they begin in $E$.  If we can show that either $op_1' = op_2'$ or that
$op_1'$ precedes $op_2'$ in $L$, then it will follow that $op_1$
precedes $op_2$ in $L$, since $op_1$ is placed immediately before
$op_1'$ and then, later, $op_2$ is placed immediately before $op_2'$.
Suppose in contradiction that $op_1'$ follows $op_2'$ in $L$.  Since
$op_1$ ends before $op_2$ begins, and $op_2$ ends before $op_2'$
begins, it follows that $op_1$ ends before $op_2'$ begins.
Furthermore, since $op_2'$ precedes $op_1'$ in $L$, when choosing
which operation in $L$ to place $op_1$ immediately before, we would
choose $op_2'$ and not $op_1'$, a contradiction.

{\em Case 3:} $op_1$ is in $B$ and $op_2$ is in $A$.  Suppose in
contradiction that $op_1$ follows $op_2$ in $L$.  Let $op_1'$ be as in
Case 2 (the earliest operation already in $L$ when $op_1$ is being
placed that starts after $op_1$ ends in $E$).  Since $op_1$ is placed
immediately before $op_1'$ in $L$, it follows that $op_1'$ also
follows $op_2$ in $L$.  By assumption, $op_2$ starts after $op_1$ ends
in $E$.  Since $op_2$ precedes $op_1'$ in $L$, when choosing which
operation in $L$ to place $op_1$ immediately before, we would choose
$op_2$ and not $op_1'$, a contradiction.

{\em Case 4:} $op_1$ is in $A$ and $op_2$ is in $B$.  By assumption,
$op_1$ ends before $op_2$ begins in $E$.  By construction, $op_2$ is
placed in $L$ immediately before $op_2'$, the earliest operation in
$L$ when $op_2$ is being placed that starts after $op_2$ ends in $E$.
Since both $op_1$ and $op_2'$ are in $A$, they are placed in $L$
according to their linearization points and thus $op_1$ precedes
$op_2'$ in $L$.  Thus $op_2$ follows $op_1$ in~$L$.
\qed
\end{proof}

\section{Algorithm for $k \ge \sqrt{2n}$}
\label{sec:kAtLeastSquareRoot}

In this section, we present a wait-free linearizable $m$-bounded $k$-multiplicative-accurate Counter, implemented using 
a shared bounded  max-register object, assuming that $k$ is a real number with $k \geq \sqrt{2n}$.
The worst-case step complexity is $O(\log \log m)$ for both the Read and Increment operations.
Pseudocode is given in Algorithm~\ref{alg:kAtLeastSquareRoot}.

\subsection{Algorithm Description}

\begin{algorithm}[htb!]
\nonl $~$
\BlankLine
  \DontPrintSemicolon
  \SetAlgoNoEnd
  \SetKwRepeat{Do}{do}{while}
  \SetKwProg{Fn}{Function}{}{end}
\nonl \textbf{Shared variable}\;
 \nonl   $\bullet$ $\mathit{logNumIncrems} :$ 
 $\lceil \log m \rceil$-bounded max register object that stores the logarithm of the number of increments exposed to the readers, initialized to $-1$.\;
 \BlankLine
\nonl \textbf{Local persistent variables}\;
\nonl $\bullet$ $\mathit{lcounter} :$  counts the number of increments invoked locally, initially $0$.\;
\nonl $\bullet$  $\mathit{threshold} :$ stores the current required number of locally-invoked increments to update 
$\mathit{logNumIncrems}$, initially $1$.\;
\nonl $\bullet$ $\mathit{nextVal} :$ stores the next value to MaxWrite into $\mathit{logNumIncrems}$, initially $0$.\;
  \BlankLine \BlankLine
\Fn{Increment()}{
    $lcounter ++$\;
    \If{$lcounter == threshold$}{
        $logNumIncrems.MaxWrite(nextVal)$ \; \label{inc:logNumIncrems}
        $nextVal ++$\;
          $lcounter \gets 0$ \;
        \lIf{$nextVal \geq 2$}{
        $threshold \gets 2 \times threshold$ 
        } 
    }
    }
  \BlankLine \BlankLine
 \Fn{Read()}{
    $r \gets logNumIncrems.MaxRead()$\; \label{read:logNumIncrems}
    \lIf{$r \ge 0$}{   
        \textbf{return} $k \cdot 2^{r}$ 
    }
    \textbf{return} $0$  \;
  }
  \BlankLine
\caption{Implementation of a $k$-multiplicative $m$-bounded counter with $k \geq \sqrt{2n}$.}
\label{alg:kAtLeastSquareRoot}
\end{algorithm}

Each process MaxWrites to a shared max-register {\it logNumIncrems} after it has experienced a certain number of Increment invocations.  
The value MaxWritten to {\it logNumIncrems} is stored in a local variable {\it nextVal}, which starts at 0 and is incremented by 1 every time {\it logNumIncrems} is written by the process.
(Of course, since {\it logNumIncrems} is a max-register, if another process has already MaxWritten a larger value, this MaxWrite will have no effect.)  
The rule for MaxWriting to {\it logNumIncrems} is that a local variable {\it lcounter} has reached a given value, stored in a local variable {\it threshold}.  Variable {\it threshold}
starts at 1 and is doubled every time {\it logNumIncrems} is MaxWritten (with the exception that the threshold remains 1 after the first MaxWrite).  Variable {\it lcounter} starts at 0, is incremented by one every time an Increment is invoked, and is reset to 0 when {\it logNumIncrems} is written.  
This procedure ensures that the value MaxWritten to {\it logNumIncrems} by the process is approximately the logarithm (base 2) of the number of Increments invoked at the process.
An exponential amount of time (and space) is saved by storing the logarithm of the number instead of the number itself, although some accuracy is lost.
Since at most $m$ Increments are assumed to occur, the max-register can be bounded by $\lceil \log m \rceil$.

During an instance of Read,
a process simply MaxReads the value $r$ of {\it logNumIncrems}, and if $r \ge 0$, then it returns $k \cdot 2^r$. 
Otherwise, the process returns $0$. 
Multiplying by $k$, which is at least $\sqrt{n}$, takes care of the uncertainty caused by the possibility of concurrent Increments.
We show that the return value falls within the approximation range 
defined by the sequential specification of the $k$-multiplicative-accurate counter.

\subsection{Proof of Linearizability}

Let $E$ be an execution of the $k$-multiplicative-accurate $m$-bounded counter implemented in Algorithm \ref{alg:kAtLeastSquareRoot}. 
We construct a linearization $L$ of $E$ by removing some specific instances of the Increment and Read
operations, then ordering the remaining operations in $E$.

Let $op$ be an incomplete Increment operation in $E$. We remove $op$ from $E$ in all but the following scenario: $op$ executes a MaxWrite on $logNumIncrems$ during $E$. We also remove from $E$, any incomplete Read operation.

From the remaining operations in $E$, we denote by $OP_w$ the set of Increment operations that do a MaxWrite on $logNumIncrems$, and $OP_l$ the set of remaining Increment operations. And let $OP_r$ denote the set of Read operations in $E$. 
We construct $L$ by first identifying linearization points in $E$ for operations in $OP_w \bigcup OP_r$ 
using Rules 1 and 2 below and then putting those operations in $L$ according to the order 
in which the linearization points occur in $E$.  
Rule 3 below describes a procedure for completing the construction of $L$ 
by inserting the operations in $OP_l$ at appropriate places.

\begin{enumerate}
    \item Each Increment operation in $OP_w$ is linearized at its MaxWrite on {\it logNumIncrems} at line \ref{inc:logNumIncrems} of Algorithm \ref{alg:kAtLeastSquareRoot}.
    \item Each Read operation in $OP_r$ is linearized at its MaxRead of {\it logNumIncrems} at line \ref{read:logNumIncrems} of Algorithm \ref{alg:kAtLeastSquareRoot}.
    \item Consider the operations in $OP_l$ in increasing order of when they begin in $E$.  Let $op$ be the next operation in $OP_l$ to be placed.  Let $op'$ be the earliest element already in $L$ such that $op$ ends before $op'$ begins and insert $op$ immediately before $op'$ in $L$. If $op'$ does not exist, then put $op$ at the end of $L$.
\end{enumerate}

Rules 1 and 2 ensure that each operation in $OP_w \cup OP_r$ is linearized at a point in the interval of its execution.
Thanks to the definition of Rule 1, Lemma~\ref{lem:generic-ordering} shows that the relative order of non-overlapping operations in the execution is preserved in the linearization:

\begin{lemma}
\label{lemma:linearizability}
Let $op_1$ and $op_2$ be two operations in $E$ such that $op_1$ ends before $op_2$ is invoked. We have that $op_1$ precedes $op_2$ in $L$.
\end{lemma}

Next, we show that the implementation respects the sequential specification of the $k$-multiplicative-accurate counter.

\begin{lemma} \label{lem:incwrite}
Each new value of $logNumIncrems$ during $E$ is an increment by $1$ of the previous value of {\it logNumIncrems}.
\end{lemma}

\begin{proof}
Let $E$ be an execution of Algorithm \ref{alg:kAtLeastSquareRoot} and consider process $p$ during $E$. 
Its local {\it nextVal} variable takes on values $0, 1, 2, 3,\ldots$.  
Thus $p$'s successive arguments used in its calls to MaxWrite() on $logNumIncrems$ are $0, 1, 2, 3, \ldots$.  Suppose in contradiction that some MaxWrite$(v)$ by $p$ causes the state of $logNumIncrems$ 
to update from its old value $u$ to the new value $v$ with $u < v-1$, causing one or more values to be skipped.  
We just argued that $p$ has previously invoked MaxWrite$(v-1)$ on $logNumIncrems$.  
Since the state of $logNumIncrems$  before $p$'s MaxWrite$(v-1)$ is always at most $u$, 
which is less than $v-1$, $p$'s MaxWrite$(v-1)$ must have taken effect 
and caused the state of $logNumIncrems$  to be at least $v-1$ 
when $p$'s MaxWrite$(v)$ is executed, a contradiction.
\qed
\end{proof}

\begin{lemma}
    Let $op$ denote an instance of the Read operation that returns $x$ in execution $E$, 
    and let $v$ be the number of Increment operations before $op$ in $L$. 
    We have $v/k \leq x \leq k \cdot v$ for $k\geq \sqrt{2n}$.
\end{lemma}

\begin{proof}
Let $r$ denote the value of {\it logNumIncrems} MaxRead during $op$ at line \ref{read:logNumIncrems} of Algorithm \ref{alg:kAtLeastSquareRoot} in $E$. Thus $x = k \cdot 2^r$.  From Lemma \ref{lem:incwrite}, the values MaxWritten to {\it logNumIncrems} in $E$ before $op$ MaxReads the value $r$ are increments of $1$, starting from $-1$ to $r$. 
Therefore, the minimum number of Increment operations necessary to reach this value of {\it logNumIncrems} is $v_{min} = 1 + \sum_{j=1}^{r} 2^{j-1}= 2^r$.
Indeed, $v+1$ Increment instances by a process are required for that process to execute MaxWrite$(v)$ on $\mathit{logNumIncrems}$ with $v \leq 1$.
Subsequently, the number of invocations required is multiplied by a factor of $2$ each time the threshold is reached. 
Thus in $E$, at least $v_{min}$ Increment instances start before $op$'s linearization point, and therefore in $L$, at least $v_{min}$ Increments occur before $op$.

Furthermore, the maximum number of Increment operations invoked in $E$
before $op$ is $v_{max} = n(1 + \sum_{i=1}^{r} 2^{i-1} )+ n(2^{r} -1)= n(2^{r+1} - 1)$. 
Each process has set $nextVal$ to $r$ (i.e., it executes $v_{min}$ Increment operations) 
plus an additional $2^{r} -1$ instances, which is the maximum number a process can count locally after setting $nextVal$ to $r$.  
Thus in $L$ at most $v_{max}$ Increments occur before $op$.

Let $v$ be the actual number of Increments that precede $op$ in $L$.  
To show that $op$'s return value $x$ is at most $k \cdot v$, observe that $x = k \cdot 2^r$, 
which equals $k \cdot v_{min}$ by the argument above, which is at most $k \cdot v$.  
To show that $x$ is at least $v/k$, note that $v\le v_{max}$, 
which equals $n(2^{r+1} - 1)$ by the argument above.  
This expression is less than $ n \cdot 2^{r+1}$, which is at most $k^2 \cdot 2^r= k \cdot x$ 
by the assumption that $k \ge \sqrt{2n}$.
\qed
\end{proof}

\subsection{Complexity Analysis}

We analyze the step complexity of Algorithm~\ref{alg:kAtLeastSquareRoot} when $logNumIncrems$ is implemented 
with the $h$-bounded max-register algorithm given by Aspnes et al.~\cite{AspnesAC2012} 
which uses 1-bit read-write registers and has a step complexity 
of $O(\log h)$ for both MaxWrite and MaxRead operations. 

\begin{lemma}
A process executes $O(\log \log m)$ steps during a call to the Read or Increment operation.
\end{lemma}
\begin{proof}
    An instance of Read calls the operation  
    MaxRead once and then computes the return value. 
    Similarly, the Increment operation calls the operation MaxWrite once
    and also computes a constant number of steps. 
    The maximum number of calls to Increment is $m$, by assumption, and thus the largest argument to MaxWrite on {\it logNumIncrems} is $\lceil \log_2 m \rceil$.  
    Since we use the $h$-bounded max register implementation from~\cite{AspnesAC2012} 
    with $O(\log h)$ step complexity, substituting $h = O(\log m)$ gives the claim.
    \qed
\end{proof}

In summary, we have:
\begin{theorem}
For any real $k \ge \sqrt{2n}$ where $n$ is the number of processes, Algorithm~\ref{alg:kAtLeastSquareRoot} is a wait-free linearizable implementation of a $k$-multiplicative-accurate $m$-bounded Counter out of read-write registers that uses $O(\log \log m)$ steps for each Read or Increment operation.
\end{theorem}

\section{Algorithm for $k >1$}
\label{sec:kGreaterThanOne}

In this section, we present a wait-free linearizable $k$-multiplicative-accurate $m$-bounded Counter working properly for any real number $k$ with $k >1$ and any positive integer $m$.
The step complexity of a Read operation is  $O(\log\log_k m)$. The step complexity of an Increment operation is $O(\max\{\log n \cdot \log(\frac{k}{k-1}n), \log \log_k m \})$.
Pseudocode is given in Algorithm~\ref{alg:kLargerThanOne}.

\subsection{Algorithm Description}

As in Algorithm~\ref{alg:kAtLeastSquareRoot}, each Increment MaxWrites the logarithm of the number of Increments into a shared max-register {\it logNumIncrems}, and each Read returns an exponential function of the value it MaxReads from {\it logNumIncrems}.  
Recall that this approach saves an exponential amount of time (and space) at the cost of the loss of some accuracy.  In order to efficiently accommodate an approximation factor of any real $k > 1$, we incorporate several new ideas.

First, the base of the logarithm is $k$, not 2, and the value returned by a Read is $k^{r+1}$, where $r$ is the value MaxRead from {\it logNumIncrems}.
As in Algorithm 1, the extra factor of $k$ accommodates the uncertainty caused by the possibility of concurrent Increments. 
The max-register is bounded by $\lceil \log_k m \rceil$.

Second, in order to estimate the number of Increments more closely, processes communicate more information through additional shared objects.
In particular, processes communicate the number of increments they have experienced through an array {\it Bucket} of ``buckets''. 
Each bucket is an exact counter, bounded by $\left\lceil \frac{kn}{k-1}\right\rceil$  
and the number of buckets is $\left\lceil \frac{(k-1)m}{n} \right\rceil$, as explained below.
To ensure that every increment in a bucket has an effect (i.e., changes the stored value), a process stops using a bucket if the stored value is larger than or equal to $\left\lceil \frac{n}{k-1}\right\rceil$.

Each process keeps a local variable {\it index}, starting at 0 and incremented by 1, that indicates the current bucket to be used. 
Each Increment increments the current bucket by one and then reads a value from that bucket.
If the bucket is ``full'', i.e., the value read is at least $n/(k-1)$, then the process moves on to the next bucket by incrementing {\it index}.
However, because of the possibility of up to $n$ concurrent Increments, the maximum value of each bucket is approximately $n/(k-1) + n = kn/(k-1)$.
Since there are at most $m$ Increments, each one causes one bucket to be incremented by one, and each bucket is incremented at least $n/(k-1)$ times, the maximum number of buckets needed is approximately $m/(n/(k-1)) = (k-1)m/n$.

Unlike in Algorithm~\ref{alg:kAtLeastSquareRoot}, the value MaxWritten to {\it logNumIncrems} in Algorithm~\ref{alg:kLargerThanOne} reflects information about the number of Increments by other processes that is learned through the buckets.  
Since each bucket holds about $n/(k-1)$ Increments, the total number of Increments is approximately $n/(k-1)$ times the number of full buckets (stored in {\it index}) plus the number of Increments read from the current, non-full, bucket.
The value written to {\it logNumIncrems} is the logarithm, base $k$, of this quantity.

\begin{algorithm}[htb!]
  \DontPrintSemicolon
 \SetAlgoNoEnd
  \SetKwProg{Fn}{Function}{}{end}
\BlankLine

\nonl \textbf{Constants:} 
$X = \frac{1}{(k-1)}$ \;
 \BlankLine   \BlankLine
\nonl \textbf{Shared variables:}\;
  \nonl      $\bullet$ $\mathit{Bucket}[\lceil  \frac{m}{Xn} \rceil]$: 
  array of $\lceil (X+1)n\rceil$-bounded  exact counter objects indexed starting at $0$,
  initialized to all $0$'s. \;
  \nonl    $\bullet$ $\mathit{logNumIncrems} :$ $\lceil \log_k m \rceil$-bounded max register object that stores the logarithm (base $k$) of the number of increments exposed to the readers, initialized to $-1$.\;
 \BlankLine   \BlankLine
\nonl  \textbf{Local persistent variables:}\;
 \nonl $\bullet$ $\mathit{index} :$ stores the current $Bucket$  index, initially $0$.\;
 \BlankLine \BlankLine
 
\Fn{Increment()}{ 
 $Bucket[index].Increment()$\; \label{incC:exactCounter}
      $val \gets Bucket[index].Read()$\;
         \label{incC:exactCounterRead}
         \If { $val < X \cdot n$ } {                   \label{incC:smallval}
      $logNumIncrems.MaxWrite(\lfloor \log_k (val + index \cdot X \cdot n) \rfloor)$ \; 
      \label{incC:maxregisterSmall} 
      }
	\Else {                   \label{incC:bigval}
      $index ++$  \;   \label{incC:incindex}
       $logNumIncrems.MaxWrite(\lfloor \log_k (index \cdot X \cdot n) \rfloor)$  \; 
       \label{incC:maxregisterBig} 
        }
} 
\BlankLine \BlankLine

  \BlankLine \BlankLine
 
 \Fn{Read()}{
    $r \gets logNumIncrems.MaxRead()$\; \label{readC:maxregister}
     \lIf {$r \geq 0$} {
     \textbf{return} $k^{r+1}$ 
     \label{readC:calcReturn}
     }
      \textbf{return} $0$\; \label{readC:case0} 
  }
  
  \BlankLine
\caption{Implementation of a $k$-multiplicative $m$-bounded counter with $k > 1$.}
\label{alg:kLargerThanOne}
\end{algorithm}

\subsection{Proof of Linearizability}

Let $E$ be an execution of the $k$-multiplicative-accurate $m$-bounded counter implemented in Algorithm \ref{alg:kLargerThanOne}. 
We construct a linearization $L$ of $E$ by removing some specific instances of the CounterIncrement
and Read operations, then ordering the remaining operations in $E$.

We remove from $E$ any incomplete Read operation. We also remove from $E$ any incomplete Increment that increments {\it Bucket}$[i]$ for some $i$ but such that there is no subsequent Read of {\it Bucket}$[i]$ in an operation that contains a MaxWrite to {\it logNumIncrems}.

From the remaining operations in $E$, we denote by $S_{CR}$: the set of (completed) Reads. We also denote by $S_{CI}^{inc}$: the set of Increments that increment {\it Bucket}$[i]$ for any $i \ge 0$; an operation in this set might or might not perform a MaxWrite on {\it logNumIncrems}.
Note that if an operation $op$ in this set is incomplete, then its bucket Increment must be followed by a Read of that bucket inside an operation that performs a 
MaxWrite; otherwise $op$ would have been excluded from consideration by the second bullet given above.

We construct $L$ by identifying linearization points in $E$ for operations in $S_{CR} \bigcup S_{CI}^{inc}$ using the following Rule 1 and Rule 2. These rules ensure that each operation in 
$S_{CR} \bigcup S_{CI}^{inc}$ is linearized at a point in its execution interval. 
$L$ is built according the following directive: when an operation $op$ has its linearization point before the linearization point of $op'$, $op$ precedes $op'$ in $L$.

\begin{enumerate}
\item  The linearization point of each operation $op \in S_{CR}$ is the enclosed
MaxRead of {\it logNumIncrems} done by $op$.
\item The linearization point of each operation $op \in S_{CI}^{inc}$
is the earlier of (i) the return of $op$ and (ii) the time of
the earliest
MaxWrite to {\it logNumIncrems} in an operation $op'$
such that $op'$ reads {\it Bucket}$[i]$ after $op$ increments 
{\it Bucket}$[i]$.
It is possible for $op'$ to be $op$.
If $op$ finishes, then at least (i) exists.
If $op$ is incomplete, then at least (ii) exists by the criteria for
dropping incomplete operations.
Operations that have the
same linearization point due to this rule appear in $L$ in any order.
\end{enumerate}

Thanks to Rules 1 and 2, the relative order of non-overlapping operations in the execution is preserved in the linearization:

\begin{lemma}
\label{lem:algC:real-time-order}
Let $op_1$ and $op_2$ be two operations in $E$ such that $op_1$ ends before $op_2$ is invoked. We have that $op_1$ precedes $op_2$ in $L$.
\end{lemma}

We call a MaxWrite to {\it logNumIncrems} {\em effective} if it changes the value of {\it logNumIncrems}.
We define $op_r$ as the Increment containing the effective MaxWrite of $r$ to {\it logNumIncrems}, for $r \ge 0$.  

The sequence of values taken on by {\it logNumIncrems} in execution $E$
(i) begins at $-1$, and
(ii) is increasing.
$op_r$ is well-defined in that there is at most one such operation for each $r$.

For each $r\geq 0$, if $op_r$ exists then let $t_r$ be the time when the effective MaxWrite of $r$ to $logNumIncrems$ occurs in $E$.

\begin{lemma}
\label{lem:algC:effective-lin-pt}
For each $r \ge 0$, if $op_r$ exists then the linearization point of $op_r$ is at or before $t_r$.
\end{lemma}

\begin{proof}
By the code, $op_r$ increments {\it Bucket}$[i]$
for some $i$, then it reads {\it Bucket}$[i]$,
and then
it performs an effective MaxWrite on {\it logNumIncrems} in $E$.
If there is no earlier
MaxWrite inside a different operation that reads {\it Bucket}$[i]$
after $op_r$ increments it,
then the linearization point of $op_r$ is its own effective MaxWrite.
\qed
\end{proof}

We next show that $L$ satisfies the sequential specification of a
$k$-multiplicative-accurate $m$-bounded counter.
We start by showing that the number of Increments
linearized before a Read cannot be too small:  specifically,
if the Read returns $x$, then the number of preceding
Increments is at least $x/k$; see Lemma~\ref{lem:algC:lower}.  
We need some preliminary lemmas.

\begin{lemma}
\label{lem:algC:no-skip}
For any process $p$, the sequence of values taken by its {\it index} variable in $E$
(i) begins at $0$, 
(ii) is increasing, and if $i\geq 1$ appears in the sequence, then so does $i-1$.
\end{lemma}

\begin{proof}
(i) is by the initialization, and (ii) is because the
$index$ variable of $p$ is only updated in line~\ref{incC:incindex} when it is incremented by one. 
\qed
\end{proof}

\begin{lemma}
\label{lem:algC:return-zero}
Let $op$ be a Read operation that returns $0$, then no Increment operation is linearized before $op$ in $L$.
\end{lemma}

\begin{proof}
We show that the first Increment by any process appears after $op$ in $L$.
Let $p$ be the process that performs $op$. Since $op$ returns $0$, $p$ MaxReads $-1$ from {\it logNumIncrems} in the execution of $op$. Thus, no MaxWrite to {\it logNumIncrems} precedes this MaxRead of {\it logNumIncrems} by $p$.

Suppose by contradiction that $op'$, the first Increment by some process, appears in $L$ before $op$. Note that $op' \in S_{CI}^{inc}$.  

Either $op'$ is linearized when it ends, or at the time of a MaxWrite to {\it logNumIncrems}. In both cases, a MaxWrite to {\it logNumIncrems} occurs in $E$ before $p$ MaxReads $-1$ from {\it logNumIncrems}. Since the argument of a MaxWrite is a value greater than or equal to $0$, we reach a contradiction.
\qed
\end{proof}

\begin{lemma}
\label{lem:algC:lower}
Let $op$ be a Read in $L$ that returns $x$
and let $y$ be the number of Increments that precede $op$
in $L$.  Then $y \ge x/k$.
\end{lemma}

\begin{proof}
{\em Case 1:} $x = 0$. By Lemma \ref{lem:algC:return-zero}, $y=0$ and the claim follows.

{\it Case 2:} $x > 0$. Then $op$ MaxReads some value $r \ge 0$ from {\it logNumIncrems}, where $x = k^{r+1}$.  
By the definition of $L$ and Lemma~\ref{lem:algC:effective-lin-pt}, $op_r$ precedes $op$ in $L$.
Let $p$ be the process that executes $op_r$. By the code, $p$ increments {\it Bucket}$[i]$, then reads {\it val}, which is at least 1, from {\it Bucket}$[i]$, and finally MaxWrites $r$ to {\it logNumIncrems}. Then {\it Bucket}$[i] \ge$ {\it val} immediately before $p$ MaxWrites to {\it logNumIncrems}. If {\it val} $< Xn$, then $r = \lfloor \log_k (val + iXn) \rfloor$, otherwise $r = \lfloor \log_k ((i+1)Xn) \rfloor$. 

According to Lemma~\ref{lem:algC:no-skip},
for each $j$, $0 \le j \le i-1$, there is an Increment instance $op_j$ performed by $p$ such that $p$ reads a value greater than or equal to $Xn$ from {\it Bucket}$[j]$ in the execution of $op_j$. By the linearization Rule 2, the linearization point of each Increment instance that applies one of the first $Xn$ increments to {\it Bucket}$[j]$ is before or at the time $op_j$ does its MaxWrite. By Lemma \ref{lem:algC:real-time-order}, $L$ respects the real-time order, and thus for each $j$, $0 \le j \le i-1$, $op_j$ precedes $op_r$ and also $op$ in $L$. Thus the number of Increments linearized before $op$ is at least $val + iXn$.  

Note that $val + iXn = k^{\log_k (val + iXn)} \ge k^{\lfloor log_k (val + iXn) \rfloor}$.  If $val < Xn$, then the exponent on $k$ is $\lfloor \log_k (val + iXn) \rfloor$.  If $val \ge Xn$, then the exponent on $k$ is at least $\lfloor \log_k((i+1)Xn) \rfloor$. In both cases, the value of the exponent on $k$ is equal to the value of $r$ for the corresponding case. Thus at least $k^r$ Increments appear in $L$ before $op$, including $op_r$.

We have $y \ge k^r = k^{r+1}/k = x/k$. The claim follows.
\qed
\end{proof}

We now show that the number of Increments linearized before
a Read cannot be too big:  specifically, if the Read
returns $x$, then the number of preceding Increments is at
most $kx$. This is proved in Lemma~\ref{lem:algC:upper}.  
We need some preliminary definitions and lemmas.

For every $i \ge 0$, let $V_i$ be the set of all Increment operations performed in $E$ such that the value of the local variable {\it index} at the beginning of the operation is $i$.

\begin{lemma}
\label{lem:algC:V-after-op}
If $op \in V_{i}$, then $op$ appears in $L$ after 
$op_r$ where $r = \lfloor \log_k(iXn) \rfloor$ for all $i \ge 1$.
\end{lemma}

\begin{proof}
We first show that $op$ begins after $t_r$ in $E$.  Let $p$ be the process executing $op$.  Since $op$ is in $V_{i}$, $p$'s {\it index} variable equals $i$ at the beginning of $op$ with $i > 0$.  By Lemma \ref{lem:algC:no-skip}, the previous value of {\it index} was $i-1$
and when $p$ incremented {\it index} to $i$, it MaxWrote $\lfloor \log_k(iXn) \rfloor$ to {\it logNumIncrems} (cf.\ Lines \ref{incC:incindex}--\ref{incC:maxregisterBig}).

Thus the first MaxWrite of $r=\lfloor \log_k(iXn) \rfloor$ to {\it logNumIncrems}, which occurs at $t_r$ by definition, precedes the beginning of $op$ in $E$. 
The  linearization point of $op$ occurs during its interval and thus follows $t_r$, which by Lemma~\ref{lem:algC:effective-lin-pt} is at or after the linearization point of $op_r$. Hence, in  $L$, $op_r$ precedes $op$.
\qed
\end{proof}

\begin{lemma}
\label{lem:algC:V-set-sizes}
$|V_i| < n(X+1)$ for all $i \ge 0$.
\end{lemma}

\begin{proof} 
We  show that {\it Bucket}$[i]$ is incremented at most $Xn + n -1$
times, for all $i \ge 0$.
After the first $Xn-1$ Increments that increment {\it Bucket}$[i]$,
each subsequent Increment by a
process $p$ reads a value greater than or equal to $Xn$ from {\it Bucket}$[i]$. Thus, $p$ moves to the next bucket by incrementing its {\it index} variable.
No subsequent Increment by $p$ contributes to $V_i$.
Thus each of the $n$ processes does at most one Increment
of {\it Bucket}$[i]$ after the first $Xn-1$ Increments of {\it Bucket}$[i]$.

Every Increment of {\it Bucket}$[i]$ by that process corresponds to one Increment (of the approximate Counter).
Since {\it Bucket}$[i]$ is incremented at most $Xn+n-1$ times,
$|V_i| < n(X+1)$.
\qed
\end{proof}

\begin{lemma}
\label{lem:algC:upper}
Let $op$ be a Read operation that returns $x$, and let $y$ be the number of Increment instances that precede $op$ in $L$.  Then $y \le kx$.
\end{lemma}

\begin{proof} 
{\it Case 1:} $x = 0$.  By Lemma \ref{lem:algC:return-zero}, no Increment operation is linearized before $op$ in $L$. Then, $y=0$ and the claim follows.

{\it Case 2:} $x > 0$.
Then $op$ MaxReads some value $r \ge 0$ from {\it logNumIncrems}, where  $x = k^{r+1}$.  Note that the MaxWrite to {\it logNumIncrems} in $op_r$ precedes $op$'s MaxRead of {\it logNumIncrems} and no MaxWrite of a value larger than $r$ precedes $op$'s MaxRead of {\it logNumIncrems}.

Also, every process has {\it index} 0 as long as {\it logNumIncrems} is less than $\lfloor \log_k(Xn) \rfloor$.  
The reason is that when a process sets its {\it index}
to a value greater than 0, it also MaxWrites  $\lfloor \log_k(Xn) \rfloor$ (or larger) to {\it logNumIncrems} (cf.\ Lines \ref{incC:incindex}--\ref{incC:maxregisterBig}). Then, we analyze the two possible cases : 

{\it Case 2.1:} $r < \lfloor \log_k(Xn) \rfloor$. Since $op_r$ MaxWrites $r = \lfloor \log_k v \rfloor$ to {\it logNumIncrems} where $v$ is the value $op_r$ read from {\it Bucket}$[0]$, we have $r \leq \log_k v < r+1$. Then, $k^r \leq v<k^{r+1}$. This means that at most $k^{r+1}-1$ Increments have been applied to {\it Bucket}$[0]$ before the Read of $op_r$.

In the following we prove that the number $y$ of Increment operation instances linearized before $op$ is
at most $\lfloor k^{r+1} \rfloor - 1$, which is less than $kx = k^{r+2}$.

Let $W$ be the set of Increment operation instances that enclose the first $\lfloor k^{r+1} \rfloor-1$   
Increments of {\it Bucket}$[0]$. Since Increments of {\it Bucket}$[0]$ are instantaneous, $W$ is well-defined.  We prove that no Increment other than those in $W$ can appear before $op$ in $L$.
Suppose by contradiction that some $op' \in S_{CI}^{inc} \setminus W$ by a process $p'$ appears in $L$ before $op$. Since $op'$ is not in $W$, and because in the execution of an Increment operation, a process first increments the {\it Bucket} and then reads its value, $p'$ reads a value $v' \ge \lfloor k^{r+1} \rfloor$ from {\it Bucket}$[0]$. 

Thus, the corresponding MaxWrite to {\it logNumIncrems} MaxWrites the value $r' = \lfloor \log_k v' \rfloor \ge r+1 > r$ to {\it logNumIncrems}. 
\begin{itemize}

\item Suppose $op'$ is linearized when it ends. Before it ends, the MaxWrite of $r' > r$ is applied to {\it logNumIncrems}. But then $op$ would read $r'$ instead of $r$ from {\it logNumIncrems}, a contradiction.

\item Suppose $op'$ is linearized at the time of the MaxWrite to {\it logNumIncrems} by some operation $op''$ that reads {\it Bucket}$[0]$ after $op'$ increments {\it Bucket}$[0]$.
Since $op'\not \in W$, the value of {\it Bucket}$[0]$ is larger than or equal to $\lfloor k^{r+1} \rfloor$.
Then the value of $op''$'s effective MaxWrite is some value $r'' \ge r' > r$. Since this MaxWrite occurs in $E$ before $op$'s MaxRead of {\it logNumIncrem}, $op$ would return $r''$ instead of $r$, a contradiction.
\end{itemize}
{\it Case 2.2:} $r \ge \lfloor \log_k(Xn) \rfloor$. 
Since $op$ precedes $op_{r+1}$ in $L$ (if $op_{r+1}$ exists),
Lemma~\ref{lem:algC:V-after-op} implies that every Increment in $V_i$ with $i\ge j$ appears after $op_{r+1}$, and thus after $op$, where $r+1 = \lfloor \log_k (jXn) \rfloor$.
Solving for $j$, we get $j \ge k^{r+1}/Xn$.
Thus $y$ is at most the number of operations in $V_0 \cup \ldots \cup V_{\lfloor k^{r+1}/Xn \rfloor - 1}$.
By Lemma~\ref{lem:algC:V-set-sizes}, 
\begin{align*}
\left| \bigcup_{j=0}^{\lfloor k^{r+1}/Xn \rfloor-1} V_j \right| 
   &\le n(X+1) \lfloor k^{r+1}/Xn \rfloor  \\
   &\le (X+1)/X \cdot Xn \lfloor k^{r+1}/Xn \rfloor \\
   &\le (X+1)/X \cdot k^{r+1} \\
   &\leq k \cdot k^{r+1} \\
   &\leq kx.
\end{align*}
\qed
\end{proof}

\subsection{Complexity Analysis}
We analyze the step complexity of Algorithm~\ref{alg:kLargerThanOne} when {\it logNumIncrems} is implemented with the $h$-bounded max register algorithm given by Aspnes et al.~\cite{AspnesAC2012} which uses 1-bit read-write registers and has a step complexity 
of $O(\log h)$ for both MaxWrite and MaxRead operations. 
Also we consider the $l$-bounded exact counter algorithm given by Aspnes et al.~\cite{AspnesAC2012} which has a step complexity of $O(\log l)$ for the Read operation and $O(\log n \cdot \log l)$ for the Increment.

\begin{lemma}
\label{lem:kGreaterThanOne}
A process executes $O(\log \log_k m)$ steps during a call to the Read, and $O(\max\{\log n \cdot \log(\frac{k}{k-1}n), \log \log_k m\})$ steps during a call to the Increment, for any $k > 1$.
\end{lemma}

\begin{proof}
    Since we use the $h$-bounded max register implementation from~\cite{AspnesAC2012} with $O(\log h)$ step complexity, substituting $h = O(\log_k m)$ gives the worst case step complexity for the MaxRead and MaxWrite operations applied to the max-register {\it logNumIncrems}.
    
    Thus, since an instance of Read does one MaxRead on {\it logNumIncrems} and then computes the return value, the claim follows for the Read operation.

    An instance of Increment does one Increment and one Read on a bucket, which is a low-level $h$-bounded exact counter where $h=\lceil (X+1)n\rceil$ and $X=\frac{1}{k-1}$.  It also does one MaxWrite on {\it logNumIncrems} whose worst-case step complexity is in $O(\log\log_k m)$.   
    
   The worst-case step complexities for the Increment and Read on the $h$-bounded exact counter are in $O(\log n \log (\lceil (X+1)\rceil) n)$ and $O(\log \lceil (X+1)\rceil n)$ with $X=\frac{1}{k-1}$, respectively. Thus, the worst case step complexity for the Increment operation is in $O(\max\{\log n \log (\frac{k}{k-1}n), \log \log_k m)\})$.
   \qed
\end{proof}

In summary, we have:
\begin{theorem}
For any real $k > 1$, Algorithm~\ref{alg:kLargerThanOne} is a wait-free linearizable implementation of a $k$-multiplicative-accurate $m$-bounded Counter out of read-write registers that uses $O(\log \log m)$ steps for each Read operation and $O(\max\{\log n \cdot \log(\frac{k}{k-1}n), \log \log_k m\})$ steps for each Increment operation.
\end{theorem}

\section{Algorithm for $k \ge 2$}
\label{sec:kAtLeastTwo}

In this section, we present an implementation of a $k$-multiplicative-accurate $m$-bounded counter for any positive integer $m$ and any integer $k \ge 2$.  
The algorithm is a variation of that in Section~\ref{sec:kGreaterThanOne}; it works for a more restricted range of $k$ but the space complexity is better, as discussed below.
The pseudocode appears in Algorithm~\ref{alg:kAtLeastTwo}. The step complexity of a Read operation is $O(\log \log_k m)$ and the step complexity of an Increment operation is $O(\max\{\log^2 n, \log\log_k m\})$.

\subsection{Algorithm Description}

When $k$ is at least 2, we can reduce the number of buckets needed exponentially, from about $(k-1)m/n$ to about $\log_k(m/n)$, while the size of each bucket at most quadruples.
The key idea is to store some of the Increments locally instead of incrementing a bucket for each one.

To deal with values of $n$ that are not powers of $k$, we use the notation $cln$ for $\lceil \log_k n \rceil$ and let $N$ be $k^{cln}$, i.e., $N$ is the smallest power of $k$ that is at least as large as $n$. 

\begin{algorithm}[htb!]

 \DontPrintSemicolon
 \SetAlgoNoEnd
  \SetKwProg{Fn}{Function}{}{end}
\BlankLine

\nonl \textbf{Constants:} 
$cln =\lceil \log_k n \rceil$ and $N = k^{cln}$ 
 \BlankLine   \BlankLine
\nonl \textbf{Shared variables:}\;
  \nonl      $\bullet$ $\mathit{Bucket}[\max\{1,\lceil \log_k \lceil \frac{m}{n} \rceil \rceil + 1\}]$:
  array of $2N$-bounded exact counter objects indexed starting at $0$, 
  initialized to all $0$'s. \;
  \nonl    $\bullet$ $\mathit{logNumIncrems} :$ $\lceil \log_k m \rceil$-bounded max register object that stores the logarithm (base $k$) of the number of increments exposed to the readers, initialized to $-1$.\;
 \BlankLine   \BlankLine
\nonl  \textbf{Local persistent variables:}\;
 \nonl $\bullet$ $\mathit{lcounter} :$ counts the number of increments invoked locally, initially $0$.\;
 \nonl $\bullet$ $\mathit{index} :$ stores the index of the current entry in $Bucket$, initially $0$.\;
 \nonl $\bullet$ $\mathit{threshold} :$ stores the current number of locally-invoked increments required to update the current entry in $Bucket$, initially $1$.\;
 \BlankLine \BlankLine
 
\Fn{Increment()} {
    $lcounter ++$\;
    \If {$lcounter == threshold$}  {
 $Bucket[index].Increment()$\; \label{inc2:exactCounter}
	     $lcounter \gets 0$ \;
      $val \gets Bucket[index].Read()$\;
         \label{inc2:exactCounterRead}
      \If{ $(index == 0)$ and $( val < N)$} { $logNumIncrems.MaxWrite(\lfloor \log_k val \rfloor)$
      \label{inc2:maxregisterZero}
      }
	\If { $val \geq N$ } {                   \label{inc2:bigval}
     $logNumIncrems.MaxWrite(cln +index)$ \; \label{inc2:maxregister}
     	$index ++$\;                         \label{inc2:incindex}
        \lIf {$index >1$} {
        $threshold \gets k \cdot threshold$
        \label{inc2:updateThreshold} }
       \lIf {$index == 1$} {
        $threshold \gets k-1 $ 
        \label{inc2:updateThresholdIndexUn}
         } } } }   \BlankLine \BlankLine

 \Fn{Read()}{
    $r \gets logNumIncrems.MaxRead()$\; \label{read2:maxregister}
     \lIf {$r \geq 0$} {
      \textbf{return} $k^{r+1}$
     \label{read2:calcReturn}
     }
      \textbf{return} $0$\; \label{read2:case0}
  }
  \BlankLine
\caption{Implementation of a $k$-multiplicative $m$-bounded counter, $k \ge 2$.}
\label{alg:kAtLeastTwo}
\end{algorithm}

Each bucket is an exact counter, bounded by $2N$ as explained below, and the number of buckets is approximately $\log_k(m/n)$.
Note that $\log_k(m/n)+1 $ buckets are enough to store $m$ Increments as the number of Increments stored in the first $i+1$ buckets is $k$ times the number stored in the first $i$ buckets, and the first bucket stores $N$ Increments.

Each process keeps a local variable, {\it index}, starting at 0 and incremented by 1.  Depending on the value of {\it index}, a certain number of Increments need to be invoked by a process before the process posts them by incrementing {\it Bucket}[{\it index}].  This number is stored in a local variable {\it threshold} and it increases as {\it index} increases.  Each process keeps track of the progress toward reaching {\it threshold} using its local variable {\it lcounter}.

When an Increment is invoked, {\it lcounter} is incremented.  If {\it threshold} has been reached, then {\it lcounter} is reset to 0, the process increments the appropriate bucket and then reads that bucket.  If the value read is less than $N$, then no further action is taken except in the corner case when {\it index} is 0, causing the logarithm (base $k$) of the value to be MaxWritten to {\it logNumIncrems}.
If the value read is at least $N$, i.e., the bucket is ``full'', then $cln + index$ is MaxWritten to {\it LogNumIncrems}, {\it index} is incremented, and {\it threshold} is updated. 
If {\it index} is 1, then the threshold is set to $k-1$, otherwise it is set to $k$ times its previous value.
The method for updating {\it threshold} is key to the correct working of the algorithm.

The reason that each bucket is bounded by $2N$ instead of $N$ is that all the processes can have the same value $i$ for their {\it index} variables when {\it Bucket}$[i]$ equals $N-1$, and then have each process start an Increment, causing $N$ additional increments to be done on {\it Bucket}$[i]$.

The Read operation is the same as in Algorithm~\ref{alg:kLargerThanOne}.

\subsection{Proof of Linearizability}

The proof of linearizability of Algorithm~\ref{alg:kAtLeastTwo} uses some of the same ideas as in the proof of linearizability for Algorithm~\ref{alg:kLargerThanOne}, although the definition of the linearization is more involved.

We start with some basic observations about the behavior of the
algorithm.  Let $E$ be any execution.
First note that each Read performs one MaxRead of {\it logNumIncrems} (at Line~\ref{read2:maxregister}), and each Increment performs at most one bucket Increment (at Line~\ref{inc2:exactCounter}), at most one bucket Read (at Line~\ref{inc2:exactCounterRead}), and at most one MaxWrite on {\it logNumIncrems} (at Line~\ref{inc2:maxregisterZero} or Line~\ref{inc2:maxregister}).

The next lemma shows that once {\it logNumIncrems} reaches $cln$, no
subsequent values are skipped\footnote{This property is not necessarily true
for smaller values, as the following counter-example shows: Let $k=2$ and $n = N
= 8$ so that $cln = 3$.  Suppose 7 processes increment {\it Bucket}$[0]$
one after the other, then they all read 7 from {\it Bucket}$[0]$ one
after the other, and then they all MaxWrite $\lfloor\log_2 7\rfloor = 2$ to
{\it logNumIncrems} one after the other.  The values 0 and 1 are
skipped.}.

\begin{lemma}
\label{lem:alg2:no-skip}
The sequence of values taken on by {\it logNumIncrems} in $E$
(i) begins at $-1$, 
(ii) is increasing, and
(iii) if $i > cln$ appears in the sequence, then so does $i-1$.
\end{lemma}

\begin{proof}
(i) is by the initialization and (ii) is because {\it logNumIncrems} is a
max-register.

(iii) Suppose in contradiction there exists a value $i > cln$ such that
$i$ is in the sequence but $i-1$ is not.
Let $p$ be the process that first MaxWrites $i$ to {\it logNumIncrems}, at some
time $t$.  By the contradiction assumption and (ii), the value of
{\it logNumIncrems} at all times before $t$ is less than $i-1$.  
Since $i > cln$,
the code ensures that $p$'s previous MaxWrite to {\it logNumIncrems} is for the
value $i-1$ (cf.\ Lines \ref{inc2:bigval}--\ref{inc2:incindex}).
But this MaxWrite to {\it logNumIncrems} occurs at some time $t' < t$
when the value of {\it logNumIncrems} is less than $i-1$.
Thus {\it logNumIncrems} is set to $i-1$ at $t'$, contradicting the assumption
that {\it logNumIncrems} never takes on the value $i-1$.  
\qed
\end{proof}

We call a MaxWrite to {\it logNumIncrems} {\em effective} if 
it changes the value of {\it logNumIncrems}.
We define $op_r$ as the Increment containing the effective 
MaxWrite of $r$ to {\it logNumIncrems}, for $r \ge 0$.  
Lemma~\ref{lem:alg2:no-skip} implies that $op_r$ is well-defined in that 
there is at most one such operation for each $r$.

We now define the linearization of $E$.
Consider all the operations in $E$ except for 
\begin{itemize}
\item any incomplete Read,
\item any incomplete Increment that does not increment a bucket 
      (i.e., that does not change any of the shared objects), and
\item any incomplete Increment that increments
      {\it Bucket}$[i]$ for some $i$ but there is no subsequent Read of
      {\it Bucket}$[i]$ in an operation that contains an effective 
      MaxWrite to {\it logNumIncrems}.
\end{itemize}

Partition the remaining operations into:
\begin{itemize}
\item $S_{CR}$: the set of (completed) Reads;
\item $S_{CI}^{inc}$: the set of Increments that 
  increment {\it Bucket}$[i]$ for any $i \ge 0$; an operation in this
  set might or might not perform a MaxWrite on {\it logNumIncrems}
  and if it does, the MaxWrite might or might not be effective.
  Note that if an operation $op$ in this set is incomplete, then its
  bucket increment must be followed by a Read of that bucket inside
  an operation that performs an effective MaxWrite; otherwise $op$ would have been excluded from consideration by the third bullet given above.
\item $S_{CI}^{silent}$: the set of (completed) Increments 
  that do not increment any bucket (and thus do not change any of the
  shared objects), i.e., the {\it silent} operations.
\end{itemize}

We construct the linearization $L$ of $S_{CR} \cup S_{CI}^{inc} \cup S_{CI}^{silent}$ in two steps.
First, we define linearization points for operations in $S_{CR} \cup S_{CI}^{inc}$ that are inside the intervals of the operations and order those operations in $L$ according to their linearization points:

\begin{itemize}
\item[{\bf L1:}] The linearization point of each operation $op \in S_{CR}$ is the enclosed MaxRead of {\it logNumIncrems} done by $op$.

\item[{\bf L2:}] The linearization point of each operation $op \in S_{CI}^{inc}$
is the earlier of (i) the return of $op$ and (ii) the time of
the earliest effective MaxWrite to {\it logNumIncrems} in an operation $op'$
such that $op'$ reads {\it Bucket}$[i]$ after $op$ increments 
{\it Bucket}$[i]$.
It is possible for $op'$ to be $op$.
If $op$ finishes, then at least (i) exists.
If $op$ is incomplete, then at least (ii) exists by the criteria for
dropping incomplete operations.
Operations that have the
same linearization point due to this rule appear in $L$ in any order.
\end{itemize}

For each $r\geq 0$, if $op_r$ exists then let $t_r$ be the time when the effective MaxWrite of $r$ to $logNumIncrems$ occurs in $E$.

\begin{lemma}
\label{lem:alg2:effective-lin-pt}
For each $r \ge 0$, if $op_r$ exists then the linearization point of $op_r$ is at or before $t_r$.
Furthermore, if $r \ge cln + 1$, then the linearization point of $op_r$ is at $t_r$.
\end{lemma}

\begin{proof}
By the code, $op_r$ increments {\it Bucket}$[i]$, where $i = 0$ if $r < cln$ 
and $i = r - cln$ otherwise, then it reads {\it Bucket}$[i]$, and then
it performs an effective MaxWrite on {\it logNumIncrems} in $E$.
If there is no earlier effective MaxWrite inside a different
operation that reads {\it Bucket}$[i]$ after $op_r$ increments it,
then the linearization point of $op_r$ is its own effective MaxWrite.
It is possible for $op_r$ to be linearized before $t_r$ when $r \le cln$.
However, as we show next, this is not possible when $r > cln$.

Now suppose $r = cln + i$, where $i \ge 1$.  Suppose the linearization point of $op_r$ is before $t_r$.  Then there is another Increment $op$ that Reads {\it Bucket}$[i]$ after $op_r$ Increments {\it Bucket}$[i]$ and MaxWrites {\it logNumIncrems} before $op_r$ does.  By the code, the value that $op$ MaxWrites is $cln + i = r$.  But this contradicts the definition of $op_r$.
\qed
\end{proof}

For the second step, we give a rule for inserting the remaining operations into $L$.  
\begin{itemize}
\item[{\bf L3:}] Consider the operations in $S_{CI}^{silent}$ in increasing
order of invocation in $E$.  Let $op$ be the next one to be placed in $L$.
Let $op'$ be the earliest operation already in $L$ such that $op$ ends
before $op'$ begins in $E$ 
and insert $op$ immediately before $op'$ in $L$.
If $op'$ does not exist, then put $op$ at the end of $L$.
\end{itemize}

By Lemma~\ref{lem:generic-ordering}, we have:

\begin{lemma}
\label{lem:alg2:real-time-order}
$L$ respects the order of non-overlapping operations in $E$.
\end{lemma}

We next show that $L$ satisfies the sequential specification of a
$k$-multiplicative-accurate $m$-bounded counter, for $k \ge 2$.
We start by showing that the number of Increments
linearized before a Read cannot be too small:  specifically,
if the Read return $x$, then the number of preceding
Increments is at least $x/k$; see Lemma~\ref{lem:alg2:lower}.  
We need some preliminary lemmas.
 
\begin{lemma}
\label{lem:alg2:lower-help-0}
If $op_{cln+1}$ exists, then at least $N$ Increments that were executed in $E$ with {\it index} variable equal to 0 appear in $L$ before $op_{cln+1}$.
\end{lemma}

\begin{proof}
Let $op$ be one of the $N$ or more Increments that increment {\it Bucket}$[0]$ before $op_{cln}$ reads {\it Bucket}$[0]$ in $E$.  The linearization point of $op$ is at $t_{cln}$ or earlier.  By definition, $t_{cln} < t_{cln+1}$.  By Lemma~\ref{lem:alg2:effective-lin-pt}, $t_{cln+1}$ is the linearization point of $op_{cln+1}$.  Thus $op$ appears in $L$ before $op_{cln+1}$.
\qed
\end{proof}

\begin{lemma}
\label{lem:alg2:lower-help}
If $op_r$ exists, then
at least $(k-1)k^{i-1} \cdot N$ Increments that were executed in $E$ with
{\it index} variable equal to $i$ appear in $L$ up to and including $op_{r}$, where $r = cln + i$ 
and $i \ge 1$.
\end{lemma}

\begin{proof}
By Lemma~\ref{lem:alg2:effective-lin-pt}, $op_r$ is linearized at $t_r$, the time of its MaxWrite to {\it logNumIncrems}.  When $op_r$ Reads {\it Bucket}$[i]$ in $E$, it gets a value at least $N$, so at least $N$ Increments on {\it Bucket}$[i]$ have already occurred in $E$, each in a distinct Increment (on the approximate Counter).
Let $op$ be one of these $N$ or more Increments other than $op_r$.
By the linearization rule, $op$ appears in $L$ before $op_r$, as it is linearized immediately before $op_r$'s MaxWrite to {\it logNumIncrems} if not earlier.

Each of the $N$ or more Increments that increment {\it Bucket}$[i]$ before $op_r$ reads it is preceded by $(k-1)k^{i-1}-1$ silent Increments by the same process, all of which are distinct.
By Lemma~\ref{lem:alg2:real-time-order}, they all appear in $L$ before
$op$ and thus before $op_r$, for a total of at least $(k-1)k^{i-1} \cdot N$.
\qed
\end{proof}

\begin{lemma}
\label{lem:alg2:lower}
Let $op$ be a Read in $L$ that returns $x$ and let $y$ be the number of Increments that precede $op$ in $L$.  Then $y \ge x/k$.
\end{lemma}

\begin{proof}
{\em Case 1:} $x = 0$.  Obviously, $y$ must be at least 0, and so $y \ge 0 = 0/k = x/k$.

{\it Case 2:} $x > 0$.  Then $op$ MaxReads some value $r \ge 0$ from {\it logNumIncrems}, where $x = k^{r+1}$.  
Recall that $op_r$ is the Increment containing the effective
MaxWrite of $r$ to {\it logNumIncrems}.
By the definition of $L$ and Lemma~\ref{lem:alg2:effective-lin-pt}, 
$op_r$ precedes $op$ in $L$.
Let $p$ be the process performing $op_r$.

{\it Case 2.1:} 
$r \le cln$.
During the execution of $op_r$, $p$'s {\it index} variable is 0 and
$r = \lfloor \log_k v \rfloor$, where $v$ is the value that $p$ reads
from {\it Bucket}$[0]$.
Thus at least $v$ increments of {\it Bucket}$[0]$ have occurred in $E$ before
the read of {\it Bucket}$[0]$ in $op_r$, 
each in a different Increment (on the approximate Counter).

Let $op'$ be one of these Increments.
Since $op' \in S_{CI}^{inc}$, its linearization point is at or before
the effective MaxWrite of $r$ to {\it logNumIncrems}, which is at or before the
linearization point of $op$.
Therefore $op'$ precedes $op$ in $L$ and $v \le y$, implying:
\begin{align*}
x &= k^{\lfloor \log_k v \rfloor + 1} \\
  &\le k^{\log_k v + 1} \\
  &= kv \\
  &\le ky.
\end{align*}
Thus $y \ge x/k$.

{\it Case 2.2:} 
$r > cln$.  Let $i = r - cln = r - \log_k N$.  
By Lemma~\ref{lem:alg2:no-skip}, the effective MaxWrite in $op_r$ is
preceded by effective MaxWrites to $r-1, r-2, \ldots, cln$ in $E$.
By Lemma~\ref{lem:alg2:effective-lin-pt}, $op_{cln}$ through $op_{r-1}$ all precede $op_r$ in $L$.
Thus by Lemmas~\ref{lem:alg2:lower-help-0} and~\ref{lem:alg2:lower-help}, the total number of Increments
appearing in $L$ up to and including $op_r$, and thus preceding $op$, 
is at least  
\begin{align*}
N + \sum_{j=1}^{i}(k-1)N \cdot k^{j-1}
    &= N + (k-1)N \left(\frac{k^i-1}{k-1}\right) \\
    &= N k^i \\
    &= k^{cln} k^{r-cln} \\
    &= k^r \\
    &= x/k.  
\end{align*}
\qed
\end{proof}

We now show that the number of Increments linearized before a Read cannot be too big:  specifically, if the Read returns $x$, then the number of preceding Increments is at most $kx$; see Lemma~\ref{lem:alg2:upper}.  
We need some preliminary definitions and lemmas.

For every $i \ge 0$, let $V_i$ be the set of all Increment operations performed in $E$ such that the value of the local variable {\it index} (belonging to the process doing the operation) at the beginning of the operation is $i$.

\begin{lemma}
\label{lem:alg2:V-after-op}
If $op \in V_{i+1}$, then $op$ appears in $L$ after $op_r$ where $r = cln + i$, for all $i \ge 0$.
\end{lemma}

\begin{proof}
We first show that $op$ begins after $t_r$ in $E$.  Let $p$ be the process executing $op$.  Since $op$ is in $V_{i+1}$, $p$'s {\it index} variable equals $i+1$ at the beginning of $op$.  By the way {\it index} is changed in the code, previously it had the value $i$ and when $p$ incremented {\it index} to $i+1$, it MaxWrote $cln+i$ to {\it logNumIncrems} (cf.\ Lines \ref{inc2:maxregister}--\ref{inc2:incindex}).  Thus the first MaxWrite of $cln+i$ to {\it logNumIncrems}, which occurs at $t_r$ by definition, precedes the beginning of $op$ in $E$.

Suppose $op \in S_{CI}^{inc}$.  Then its linearization point occurs during its interval and thus follows $t_r$, which, by Lemma~\ref{lem:alg2:effective-lin-pt}, is at or after the linearization point of $op_r$.  So $op$ follows $op_r$ in $L$.

Suppose $op \in S_{CI}^{silent}$.  If it is placed at the end of $L$, then this is after $op_r$.
Suppose $op$ is placed in $L$ immediately before operation $op'$ already in $L$ that begins after $op$ ends in $E$.
Recall that silent operations are considered for placement in $L$ in increasing order of invocation.
Since $op'$ begins after $op$ begins in $E$ but is already in $L$, $op'$ cannot be silent.
Since $op'$ is not silent, its linearization point is inside its interval of execution.
As argued at the beginning of the proof, $op$ starts after $t_r$ in $E$, and thus the linearization point of $op'$ is after that of $op_r$.  Therefore, $op_r$ appears in $L$ before $op'$, and $op$ appears in $L$ between $op_r$ and $op'$, i.e., after $op_r$.
\qed
\end{proof}

\begin{lemma}
\label{lem:alg2:V-set-sizes}
$|V_0| \le (2N-1)$
and
$|V_i| \le (2N-1)(k-1)k^{i-1}$ for all $i \ge 1$.
\end{lemma}

\begin{proof} 
We first show that {\it Bucket}$[i]$ is incremented at most $2N-1$ times, for all $i \ge 0$.
After the first $N-1$ Increments that increment {\it Bucket}$[i]$, each subsequent Increment, say by process $p$, reads a value at least $N$ from {\it Bucket}$[i]$, causing $p$ to move to the next bucket by incrementing its {\it index} variable.
No subsequent Increment by $p$ contributes to $V_i$.
Thus each of the $N$ processes does at most one Increment of {\it Bucket}$[i]$ after the first $N-1$ Increments of {\it Bucket}$[i]$.

When the {\it index} variable of a process is 0, every Increment of {\it Bucket}$[0]$ by that process corresponds to one Increment (of the approximate Counter).
Since {\it Bucket}$[0]$ is incremented at most $2N-1$ times, $|V_0| \le 2N-1$.

Now suppose $i \ge 1$.
Every time a process increments {\it Bucket}$[i]$ as part of an Increment (of the approximate Counter), it has previously done $\theta-1$ silent Increments with {\it index} equal to $i$, where $\theta$ is the value of $p$'s {\it threshold} variable.
By the code, $\theta = (k-1)k^{i-1}$.
Thus each Increment of {\it Bucket}$[i]$ contributes $(k-1)k^{i-1}$ Increments to $V_i$.
Therefore the total number of elements in $V_i$ is at most $(2N-1)(k-1)k^{i-1}$.
\qed
\end{proof}

\begin{lemma}
\label{lem:alg2:upper}
Let $op$ be a Read in $L$ that returns $x$ and let $y$ be the number of Increments that precede $op$ in $L$.  Then $y \le kx$.
\end{lemma}

\begin{proof} 
{\it Case 1:} $x = 0$.  We show that $y = 0$ and thus $y = 0 \le k \cdot 0 
= kx$.
Since $op$ returns 0, it MaxReads $-1$ from {\it logNumIncrems} in $E$.
Thus no MaxWrite to {\it logNumIncrems} precedes $op$'s MaxRead of
{\it logNumIncrems} in $E$.

We will show that the first Increment by any process appears in $L$ after $op$; then Lemma~\ref{lem:alg2:real-time-order} will imply that no Increment appears in $L$ before $op$.
Suppose in contradiction that $op'$, the first Increment by some process, appears in $L$ before $op$.  Note that $op' \in S_{CI}^{inc}$.  
If $op'$ is linearized when it ends, then its MaxWrite to {\it logNumIncrems} occurs in $E$ before $op$'s MaxRead of {\it logNumIncrems}, a contradiction.
If $op'$ is linearized at the time of an effective MaxWrite to {\it logNumIncrems}, then this MaxWrite occurs in $E$ before $op$'s MaxRead of {\it logNumIncrems}, a contradiction.

{\it Case 2:} $x > 0$.
Then $op$ MaxReads some value $r \ge 0$ from {\it logNumIncrems}, where 
$x = k^{r+1}$.  Note that in $E$, the MaxWrite to {\it logNumIncrems}
in $op_r$ precedes $op$'s MaxRead of {\it logNumIncrems} and no MaxWrite
of a value larger than $r$ precedes $op$'s MaxRead of {\it logNumIncrems}.
Thus there is no effective MaxWrite in between $op_r$'s MaxWrite to and
$op$'s MaxRead of {\it logNumIncrems}.

{\it Case 2.1:} $r < cln$.  First note that, in $E$, every process has {\it index} 0 as long as {\it logNumIncrems} is less than $cln$.  The reason is that when a process sets its {\it index} to a value greater than 0, it also MaxWrites a value at least $cln$ to {\it logNumIncrems} (cf.\ Lines \ref{inc2:maxregister}--\ref{inc2:incindex}).

Since $op_r$ writes $r = \lfloor \log_k v \rfloor$ to {\it logNumIncrems}
where $v$ is the value it reads from {\it Bucket}$[0]$, $v$ can be as small
as $k^r$ and as large as $k^{r+1} - 1$.  In order to maximize the upper bound, assume $v = k^{r+1} - 1$.
Let $W$ be the set of Increments that enclose the first
$k^{r+1} - 1$ increments of {\it Bucket}$[0]$.
All the elements of $W$ may appear before $op$ in $L$.
Suppose for contradiction that some $op' \in S_{CI}^{inc} \setminus W$
appears in $L$ before $op$.
Since $op'$ is not in $W$, it reads some $v' \ge k^{r+1}$ from 
{\it Bucket}$[0]$ and thus it MaxWrites $r' = \lfloor \log_k v' \rfloor \ge
r+1 > r$ to {\it logNumIncrems}.

Suppose $op'$ is linearized when it ends.  Before it ends, it MaxWrites $r' > r$ to {\it logNumIncrems} in $E$.  But then $op$ would MaxRead $r'$ instead of $r$ from {\it logNumIncrems}, a contradiction.

Suppose $op'$ is linearized at the time of the effective MaxWrite to {\it logNumIncrems} by some operation $op''$ that Reads {\it Bucket}$[0]$ after $op'$ Increments {\it Bucket}$[0]$.
Then the value of $op''$'s effective MaxWrite is some value $r'' \ge r' > r$.
Since this effective MaxWrite occurs in $E$ before $op$'s MaxRead of {\it logNumIncrem}, $op$ would return $r''$ instead of $r$, a contradiction.

Since every process has {\it index} equal to 0 up to $op$, no silent Increments can start before then and thus none can appear in $L$ before $op$.

Thus $y \le k^{r+1} - 1 < kx$.

{\it Case 2.2:} $r \ge cln$. 
By Lemma~\ref{lem:alg2:V-after-op}, no operation in $V_j$ for any $j \ge i+2$
can appear before $op$ in $L$, where $r = cln + i$.
The reason is that $op$ precedes $op_{r+1}$ in $L$, if it exists. 
By Lemma~\ref{lem:alg2:V-set-sizes}, 

\begin{align*}
\left| \bigcup_{j=0}^{i+1} V_j \right| 
    &\le (2N-1) + \sum_{j=1}^{i+1} (2N-1) \cdot (k-1)k^{j-1} \\
    &= (2N-1) + (2N-1)(k-1)\left(\frac{k^{i+1}-1}{k-1}\right) \\
    &= (2N-1) k^{i+1} \\
    &= (2N-1) k^{r - cln + 1} \\
    &= \left(2 - \frac{1}{N}\right) k^{r+1} \\
    &\le kx \mbox{\qquad since $k \ge 2$.}  
\end{align*}
\qed
\end{proof}

By Lemmas~\ref{lem:alg2:real-time-order},~\ref{lem:alg2:lower}, and~\ref{lem:alg2:upper}, the algorithm is a linearizable implementation of a $k$-multiplicative-accurate $m$-bounded counter.

\subsection{Complexity Analysis}

Let $m$ be the maximum number of Increment operations invoked in any execution of Algorithm~\ref{alg:kAtLeastTwo}.  
Then the largest value ever MaxWritten to {\it logNumIncrems} is $\lceil \log_k m \rceil$.  We implement {\it logNumIncrems} with the algorithm in \cite{AspnesAC2012} for an $h$-bounded max-register, where $h = \lceil \log_k m \rceil$.  Thus the number of steps for both MaxRead and MaxWrite on {\it logNumIncrems} is $O(\log h) = O(\log \log_k m)$.

Each bucket is incremented at most $2N$ times.  We implement each bucket using the algorithm in \cite{AspnesAC2012} for an exact $2N$-bounded counter.  
The number of steps for incrementing a bucket is $O(\log n \cdot \log(2N)$, which is $O(\log^2 n)$.
The number of steps for reading a bucket is $O(\log(2N)) = O(\log n)$.

Thus the number of steps required for a Read, which performs a MaxRead on {\it logNumIncrems}, is $O(\log \log_k m)$.
Each Increment consists of at most one Increment of a bucket, at most one Read of a bucket, and at most one MaxWrite of {\it logNumIncrems}, requiring 
$O(\log^2 n + \log \log_k m) = O(\max\{\log^2 n, \log \log_k m\})$ steps.

Algorithm~\ref{alg:kAtLeastTwo}, which requires $k \ge 2$, and Algorithm~\ref{alg:kLargerThanOne}, which works for any $k > 1$, both use a shared max-register that is bounded by $\lceil \log_k m \rceil$.
They both also use an array of shared buckets.  In Algorithm~\ref{alg:kAtLeastTwo}, the number of buckets is, roughly, $\log_k (m/n)$, while each bucket is, roughly, $(n+n)$-bounded.
In contrast, Algorithm~\ref{alg:kLargerThanOne} uses $(k-1)m/n$ buckets, each bounded by $n+n/(k-1)$.
When $k \ge 2$, the number of buckets used by Algorithm~\ref{alg:kAtLeastTwo} is exponentially smaller than that used by Algorithm~\ref{alg:kAtLeastTwo} and the bound on each bucket in Algorithm~\ref{alg:kAtLeastTwo} is less than four times
that needed in Algorithm~\ref{alg:kLargerThanOne}.

In summary, we have:
\begin{theorem}
For any integer $k \ge 2$, Algorithm~\ref{alg:kAtLeastTwo} is a wait-free linearizable implementation of a $k$-multiplicative-accurate $m$-bounded Counter out of read-write registers that uses 
$O(\log \log m)$ steps for each Read operation and 
$O(\max\{\log^2 n,$ $\log \log_k m\})$ steps for each Increment operation.
\end{theorem}

\section{Conclusion}

We have presented three algorithms for implementing a wait-free linearizable $k$-multiplicative-accurate $m$-bounded counter using read-write registers.  By combining the assumption that the number of increments is bounded by $m$ and relaxing the semantics of the object to allow a multiplicative error of $k$, our algorithms achieve improved worst-case step complexity for the operations over prior algorithms for exact or unbounded counters.  The three algorithms provide different tradeoffs between the step complexity, space complexity, and range of values for $k$, as depicted in Table~\ref{table:3-algs}.

A natural open question is to find the optimal worst-case step complexity of {\it CounterIncrement} for such implementations and see if it can be achieved simultaneously with an optimal {\it CounterRead}.  Similarly, finding a tight bound on the space complexity would be interesting.  Would allowing the implementation to use more powerful primitives, such as the historyless ones considered in~\cite{JayantiTT2000}, help?  Finally, we are not aware of any {\it worst-case} results for {\it approximate} counters in the {\it unbounded} case or of any {\it amortized} results for {\it exact} counters in the {\it bounded} case; such results would flesh out our general understanding of counter implementations.

\bibliographystyle{splncs04}
\bibliography{biblio}

\end{document}